\documentclass[a4paper]{article}

\usepackage{INTERSPEECH2019}
\usepackage{multirow}

\title{Toward Cross-Domain Speech Recognition with End-to-End Models}
\name{Thai-Son Nguyen, Sebastian St\"uker, Alex Waibel}
\address{Institute for Anthropomatics and Robotics, Karlsruhe Institute of Technology}
\email{thai.nguyen@kit.edu}

\begin{document}

\maketitle
\begin{abstract}
In the area of multi-domain speech recognition, research in the past focused on hybrid acoustic models to build cross-domain and domain-invariant speech recognition systems. In this paper, we empirically examine the difference in behavior between hybrid acoustic models and neural end-to-end systems when mixing acoustic training data from several domains. For these experiments we composed a multi-domain dataset from public sources, with the different domains in the corpus covering a wide variety of topics and acoustic conditions such as telephone conversations, lectures, read speech and broadcast news. We show that for the hybrid models, supplying additional training data from other domains with mismatched acoustic conditions does not increase the performance on specific domains. However, our end-to-end models optimized with sequence-based criterion generalize better than the hybrid models on diverse domains. In term of word-error-rate performance, our experimental acoustic-to-word and attention-based models trained on multi-domain dataset reach the performance of domain-specific long short-term memory (LSTM) hybrid models, thus resulting in multi-domain speech recognition systems that do not suffer in performance over domain specific ones.
Moreover, the use of neural end-to-end models eliminates the need of domain-adapted language models during recognition, which is a great advantage when the input domain is unknown.
% ThaiSon: modified to be not over 200 words.
\end{abstract}


\section{Introduction}
Automatic speech recognition (ASR) systems based on a naive Bayes classifier have been widely studied in the past. Traditionally, given a pre-processed audio signal, i.e. the feature vector sequence $X$, these systems search for the word sequence $W$ with the highest posterior probability $P(W\mid X)$ by maximizing the product $P(X\mid W)P(W)$. The class conditional probability $P(X\mid W)$ is called the acoustic model and is realized with Hidden Markov Models (HMMs) \cite{rabiner89}. The prior probability $P(W)$ is called the language model and generally computed using n-gram language models \cite{ney1995}.

While in the past the emission probabilities in the HMMs of the acoustic model were calculated using Gaussian Mixture Models, lately the use of artificial neural networks (ANNs) has given significant performance boosts. Many different kinds of ANNs have been studied for this purpose, e.g., deep feed-forward networks, time-delay neural networks (TDNN) \cite{waibel1989tdnn}, or long short-term memory (LSTM) \cite{hochreiter1997long} networks. These hybrid HMM/ANN systems currently achieve state-of-the-art performance on many tasks.

However, experience has shown that these systems usually work best when trained for a specific domain on domain specific data. This is true for both, the acoustic model and the language model. When mixing training data from multiple domains, performance may decrease on specific domains compared to models that were only trained on in-domain data.

Lately new types of recognition systems have been developed that directly map the acoustic feature vector sequence $X$ to a symbol sequence by utilizing neural networks, without using the naive Bayes classifier network, that is without the distinction between the two models, the acoustic model and the language model. These systems perform the mapping either in a sequence-labeling fashion, or, alternatively, theses systems work in a sequence-to-sequence classification fashion. The symbols in the output sequence can be phones, letters or directly words. In the case of letters and phones as output symbols, further processing, such as a squashing function or a decoding on top of the output symbol sequence is necessary. These so called end-to-end systems have recently come close in performance to the hybrid systems, sometimes matching performance.

In this paper we will show, that the performance of these end-to-end systems, unlike with hybrid systems, even improves on individual domains when trained in a multi-domain fashion compared to the performance of their domain specific variants. We do so by systematically examining the behaviour of a variety of hybrid and end-to-end models on a multi-domain training set, that we composed several freely-available corpora that feature a variety of different topics and acoustic conditions, such as telephone conversations, lectures, read speech and broadcast news.

\section{Prior Work}
Previous studies on multi-domain and domain-invariant speech recognition can be roughly divided into two general approaches.
The first approach focuses on exploiting additional speech data to train acoustic models which then become invariant to specific acoustic conditions, resulting in domain-invariant speech recognition systems. E.g., \cite{yu2013feature,kim2017generation} used simulated noisy utterances together with clean training data to achieve invariance to background noise. Similar to that, the authors in \cite{yu2013feature} used a mixed bandwidth training dataset to help the acoustic model generalize to multiple sampling rates. In \cite{peddinti2016far}, far-field speech recognition is significantly improved by exploiting large-scale simulated data for training deep neural networks. \cite{narayanan2018toward} build a multi-domain speech recognition system by pooling a huge amount of training data from several sources and simulated conditions like background noise, codecs and sample rates.

The second general approach utilizes adaptation techniques to adapt speech recognition models trained on one domain with a lot of data to other domain with limited data. In \cite{li2017large}, the authors utilize a student-teacher framework to adapt from clean to noisy environments and from adults to children. The transfer learning in \cite{ghahremani2017investigation} is used to adapt a neural network model trained on Switchboard \cite{godfrey1992switchboard} to different domains with smaller training sizes.

So far, most studies on multi-domain speech recognition are based on conventional hybrid acoustic models. \cite{mirsamadi2017multi} recently tried to understand how a recurrent neural network with letter-based CTC criterion learns to produce domain-invariant features from a small meeting dataset with different far-field conditions.

\section{Multi-domain Dataset}
\label{sec:dataset}

\subsection{Dataset}
Several public speech corpora have been released for speech recognition research. However, to the best of our knowledge, there has not been such a common corpus for the study of multi-domain speech recognition. We composed a multi-domain training dataset which consists of four well-known corpora: Switchboard~\cite{godfrey1992switchboard}, TED-LIUM~\cite{hernandez2018ted}, Libri Speech~\cite{panayotov2015librispeech} and Hub4 (LDC97S44 \& LDC98S71). The statistics of the multi-domain set are described in Table \ref{tab:dataset}. Basically, it includes 1,282 hours of speech data coming from different domains such as \textit{telephone conversations}, \textit{talks and lectures}, \textit{read speech} and \textit{broadcast news}. Compared to an internal multi-domain set reported in \cite{narayanan2018toward}, our composed multi-domain set is fairly distributed in which there is no disproportionally large sub-set.

To evaluate the performance of our models, we use the Hub5'00 test set (LDC2002S09), the TED-LIUM test set, the Libri test-clean set and the Hub4 eval set (LDC97S66) as the evaluation sets for the corresponding domains.

Since the audio from Switchboard and Hub5'00 corpora was originally sampled at 8kHz, we converted it to 16kHz to have an uniform format. We do not reconstruct missing bands since neural network training can handle multiple sample rates simultaneously as shown in \cite{yu2013feature}.

\begin{table}[t]
	\caption{The training and test datasets for cross-domain speech recognition.}
	\label{tab:dataset}
	\vspace{-0.1cm}	
	\setlength{\tabcolsep}{1pt}
	\centering
	\begin{tabular}{lcccc}
		\toprule
		\textbf{Dataset} & \textbf{Domain} & \textbf{Hours} & \textbf{\#Utt} \\
		\midrule
		Switchboard    & Telephone Conversation & 318 & 263K \\
		TED-LIUM       & Lecture Presentation & 453 & 268K \\
		Libri          & Read Speech & 363 & 104K \\
		Hub4           & Broadcast News & 148 & 125K \\
		\midrule
		Multi-Set      & Multiple  & 1282 & 760K \\
		\midrule
		Hub5-2000 (Hub5'00) & Telephone Conversation & 3.79 & 4458 \\
		TED-LIUM test (TED) & Lecture Presentation & 2.61 & 1155 \\
		Libri test (Libri)  & Read Speech & 5.4 & 2620 \\
		Hub4 eval (Eval98)  & Broadcast News & 2.81 & 825 \\	
		\bottomrule
	\end{tabular}
	\vspace{-0.0cm}
\end{table}

\subsection{Domain Analysis}
\label{sec:domain_separation}

First we describe the difference in the domains of the speech corpora through long-term and short-term characteristics. The long-term characteristics refer to speaker accents or speaking styles while the short-term characteristics are more related to acoustic conditions such as microphone, background noises, sample rate, codec, etc. While long-term characteristics can be efficiently identified by humans, short-term characteristics are more difficult to measure and describe. To better understand the acoustic similarities of the speech corpora in the multi-domain dataset, we have performed the following experiment.

We trained a classification model to discriminate among the speech frames from different corpora. The classifier is implemented as a feed-forward neural network with 3 layers of 500 units, and with a softmax layer for 4 different domains. Supplying a window of several consecutive frames as input, the network model learns to classify domain labels. When using a window size of 11 frames, the network model trained well on \textit{Multi-Set} and eventually achieved an accuracy of 92.6\% on the heldout set. We achieved an accuracy of 96.5\% when increasing the window size up to 31 frames. This gives an initial idea that the 4 speech corpora used in the multi-domain set are easy to distinguish by short-term acoustic features.

To further explore the similarities between the specific domains, we used the heldout data to generate a confusion matrix as shown in Table \ref{tab:confusion}. The confusion matrix is presented with the percentages of correct and incorrect predictions that the domain classifier assigns for the samples of the same ground-truth domain labels.

As can be observed, the domain classifier can detect \textit{Switchboard} with very high accuracy (probably because of missing bands), followed by \textit{Libri}. It has more confusion when detecting \textit{TED-LIUM} and \textit{Hub4}, roughly indicating that these two domains are closer.
%while \textit{Switchboard} is very different from others.

\begin{table}[t]
	\caption{The confusion matrix for the similarities between the domains. The columns of the table indicate the groups for different ground-truth domain labels.}
	\label{tab:confusion}
	\vspace{-0.1cm}	
	\setlength{\tabcolsep}{3pt}
	\centering
	\begin{tabular}{lcccc}
		\toprule
		& \textbf{Switchboard} & \textbf{TED-LIUM} & \textbf{Libri} & \textbf{Hub4} \\
		\midrule
		Switchboard   & 0.99945  & 0.00015  & 0.00008  & 0.00072 \\
		TED-LIUM      & 0.00016  & 0.95965  & 0.02425  & 0.06569 \\
		Libri         & 0.00006  & 0.01985  & 0.96635  & 0.03500 \\
		Hub4          & 0.00033  & 0.02035  & 0.00932  & 0.89859 \\
		\bottomrule
	\end{tabular}
	\vspace{-0.2cm}
\end{table}

\section{Hybrid Acoustic Model}
Conventional speech recognition systems with artificial neural network (ANN) based acoustic models using the hybrid Hidden Markov Models (HMM) / ANN approach \cite{bourlard2012connectionist,robinson1994application} have achieved state-of-the-art performance and are widely used in many research and application areas. In this section, we discover the abilities as well as the limitations of these models when performing on the multi-domain dataset.

\subsection{Setup of Hybrid models}
From a bootstrap system built on a part of \textit{Multi-Set}, we used a common cluster-tree of 8,000 context-dependent phonemes and the same forced-alignment system to provide frame-based labels for all domain-specific and multi-domain training sets. We then trained hybrid acoustic models using both a feed-forward neural network (FFNN) and a long short-term memory (LSTM) network. The FFNN models consist of 7 layers of 2,000 units while bidirectional LSTM models have 5 layers with 320 units each. 40 log mel filter-bank features which are mean and variance normalized per utterance are used for all models. For FFNNs, we used a window of 15 consecutive frames as the input, while for LSTMs, we generated sub-sequences of 50 frames with a moving step of 25 frames from the training utterances.

The hybrid acoustic models were decoded with domain-adapted language models (LM) for individual test sets. Specifically, the LM for \textit{Hub5'00} was built from the transcripts of the Switchboard and Fisher corpora, while the standard LM for \textit{Libri} is described in \cite{panayotov2015librispeech}. We used the same Cantab LM \cite{williams2015scaling} for \textit{TED} and \textit{Hub4}. To investigate the influence of the domain-adapted language models on recognition performance, we used an additional LM which was built from the transcripts of the \textit{Multi-Set} set. 

\subsection{Results of Hybrid models}
\label{sec:hybrid_sys}

In the first 4 rows of Table \ref{tab:hybrid_sys}, we present the WER performance of the hybrid acoustic models trained on individual domain-specific training sets and evaluated with all the test sets. As can be observed, the domain-specific models only perform well on the test sets that match the training domain conditions, and can be very poor on out-of-domain test sets. On \textit{Eval98} the recognizer trained on an in-domain dataset with much smaller size still outperforms other training sets. At the worst case of mismatching, the recognition performance hugely drops shown on \textit{TED}. These observations substantiate the importance of in-domain data in building hybrid speech recognition.
The cross comparisons also reveal the similarities between the individual training sets. For example, \textit{Switchboard} and \textit{Libri} are very different from the others while \textit{TED-LIUM} and \textit{Hub4} are closer corpora. These results are consistent with the analysis in Section \ref{sec:domain_separation}.

We evaluated the multi-domain model with all the test sets as in the last three rows of Table \ref{tab:hybrid_sys}. The WER performance of the multi-domain model shows two interesting facts for the combination of speech corpora of different domains. First, when two training corpora are close enough (e.g. \textit{TED-LIUM} and \textit{Hub4}), they can supplement each other so that the hybrid acoustic model can benefit from the mixed data training. Second, for the case of \textit{Switchboard} and \textit{Libri}, the recognition performance is hardly improved when the additional training datasets are diverse.

These results also reveal the abilities as well as the limitations of the hybrid speech recognition approach. On one hand, the hybrid models are capable of modeling short windows of frames from a mixed domain dataset and actually produce no performance loss in comparison to specific domain modeling. However, on the other hand, when acoustic conditions are too diverse, the hybrid models cannot generalize well which show their limitations in exploiting multi-domain speech data.

The other limitation of a conventional hybrid model is that it always requires a domain-adapted language model for inference. We investigated the influence of domain-adapted language models by decoding the multi-domain model with two different LMs. As can be seen, the performance of the multi-domain model clearly degrades when switching to the \textit{Multi-Set} LM which partly includes in-domain data, and largely drops for the Cantab LM which does not match the domains of the test sets.

\begin{table}[t]
	\caption{The WER performance of hybrid systems with FFNN and LSTM (in brackets) acoustic models. The columns of the table indicate the different test sets while the rows show the used training sets.}
	\label{tab:hybrid_sys}
	\vspace{-0.1cm}	
	\setlength{\tabcolsep}{1.0pt}
	\centering
	\begin{tabular}{lcccc}
		\toprule
		& \textbf{Hub5'00} & \textbf{TED} & \textbf{Libri} & \textbf{Eval98} \\
		\midrule
		Switchboard   & \textbf{23.3 (18.3)} & 65.7 & 22.5 & 63.1 \\
		TED-LIUM      & 54.3 & \textbf{12.0 (10.5)} & 8.1 & 17.6\\
		Libri         & 61.6 & 18.8 & \textbf{6.5 (5.9)} & 22.0\\
		Hub4          & 37.1 & 15.0 & 9.7 & \textbf{14.5 (12.8)} \\
		\midrule
		Multi-Set     & \textbf{23.2 (18.3)} & \textbf{11.1 (9.6)} & \textbf{6.4 (5.8)} & \textbf{13.6 (11.6)} \\
		\;+Multi-Set LM\; & 24.2 (19.2) & 12.3 (11.1) & 10.1 (8.4) & 14.0 (11.8) \\
		\;+Cantab LM & 27.5 (22.5) & - & 10.7 (8.6) & - \\
		\bottomrule
	\end{tabular}
	\vspace{-0.2cm}
\end{table}

\section{End-to-End Models}
\label{sec:e2e_sys}
The end-to-end speech recognition models have received significant interest in recent years due to the simplification of the recognition process by using a single neural network to estimate the direct mapping from acoustic signals to textual transcription. One of the biggest advantages of end-to-end systems is to learn both acoustic and language modeling with a unified neural network in a single training process, which then avoids the need of language models for inference as in conventional hybrid systems. This advantage is very attractive for the application of multi-domain speech recognition in which the domain of input speech is unknown or difficult to determine.

In this paper, we investigate the construction of end-to-end systems for the cross-domain speech recognition task. The two investigated end-to-end systems, acoustic-to-word \cite{soltau2016neural} and attention-based models \cite{chorowski2015attention}, do not require additional language models during inference.

\subsection{Acoustic-to-word}
\label{sec:a2w}
The acoustic-to-word (A2W) model based on the Connectionist temporal classification (CTC)~\cite{graves2006connectionist} criterion was first introduced in \cite{sak2015fast} as a natural end-to-end model directly targeting words as output. In \cite{soltau2016neural}, the authors have successfully built a direct A2W system that achieves state-of-the-art speech recognition performance by leveraging 125,000 hours of training data collected from Youtube videos. Later on, \cite{audhkhasi2017direct,audhkhasi2018building} proposed training optimization to train A2W models on the standard Switchboard 300 hours training set which results in competitive performance with other end-to-end approaches.

One of the major difficulties of training A2W system is the data sparsity problem. While \cite{soltau2016neural} has alleviated this problem by using exceptionally large training data, \cite{audhkhasi2017direct,audhkhasi2018building} have used pre-trained CTC-phone models and and used GloVe word embeddings~\cite{pennington2014glove} to initialize acoustic-to-word models on a moderately sized training data. We use the multi-task training approach proposed in \cite{nguyen2019shareencode} to directly train A2W models on the domain-specific and cross-domain data sets.

Specifically, to build A2W models we use 5 LSTM layers of 320 units. We also keep the same feature extraction as for the hybrid acoustic models. For each training set, we find words appearing more than 5 times in the transcripts to build target units for the corresponding A2W model. The second task of the multi-task network is always the framewise classification of 8000 context-dependent phonemes. We adopted a down-sampling on acoustic features performed by stacking two consecutive frames followed by the drop of one frame. Stochastic gradient descent (SGD) with New-bob training schedule are used for model optimization. Initial learning rates are set as high as possible for individual training, and then is decayed by a factor of 0.8 after 12 epochs. 

\subsection{Sequence-to-sequence}
\label{sec:seq2seq}
Sequence-to-sequence attention-based speech recognition models \cite{chorowski2015attention,bahdanau2016end,chan2016listen} use a single neural network that consists of an encoder recurrent neural network (RNN) and a decoder RNN, and uses an attention mechanism to connect between them. The decoder is analogous to a language model due to attention-based model being trained to provide a probability distribution over sequences of labels (words or characters). The encoder converting low level acoustic features into higher level representation is analogous to the RNN of an CTC model.

We follow the approach in \cite{nguyen2019shareencode} in which we took the pre-trained LSTM layers from the A2W network (trained with the same data set) to initialize the encoder of attention-based models. For the decoder, we used only one uni-directional LSTM layer. Adam \cite{kingma2014adam} and New-bob schedules are used to optimize the attention-based models. We experimented with sequence-to-sequence models using characters and words as target units. For both label units, we use a beam search with the beam size of 12 for decoding.

\subsection{Results of end-to-end models}
\begin{table}[t]
	\caption{The performance of acoustic-to-word and sequence-to-sequence models trained on domain-specific and multi-domain data sets.}
	\label{tab:end2end_sys}
	\vspace{-0.1cm}	
	\setlength{\tabcolsep}{4pt}
	\centering
	\begin{tabular}{lcccc}
		\toprule
		& \textbf{Hub5'00} & \textbf{TED} & \textbf{Libri} & \textbf{Eval98} \\
		\midrule
		\multicolumn{5}{l}{\textit{Char Seq2Seq Model}} \\
		\;\; Switchboard   & \textbf{22.9} & 45.4 & 35.5 & *60.2 \\
		\;\; TED-LIUM      & 60.1 & \textbf{13.0} & 17.0 & *29.0 \\
		\;\; Libri         & 72.7 & 34.3 & \textbf{10.3} & *52.1 \\
		\;\; Hub4          & 42.2 & 25.7 & 23.8 & \textbf{*25.5} \\
		\;\; Multi-Set   & \textbf{18.2} & \textbf{10.6} & \textbf{7.6} & \textbf{*20.8} \\
		\multicolumn{5}{l}{\textit{Word Seq2Seq Model}} \\
		\;\; Domain-Specific & 22.4 & 12.8 & 11.2 & 23.9 \\
		\;\; Multi-Set   & \textbf{18.5} & \textbf{10.6} & \textbf{8.5} & \textbf{11.9} \\
		\multicolumn{5}{l}{\textit{Acoustic-to-word Model}} \\
		\;\; Domain-Specific & 23.8 & 14.2 & 12.2 & 19.4 \\
		\;\; Multi-Set   & \textbf{19.4} & \textbf{11.3} & \textbf{8.9} & \textbf{11.9} \\		
		\bottomrule
	\end{tabular}
	\vspace{-0.2cm}
\end{table}

As done for the hybrid acoustic systems, we trained individual end-to-end models for different domain-specific and multi-domain training sets and evaluated them with the proposed test sets. For character-based models, we used an unified label set of 52 characters while the word-based sequence-to-sequence and acoustic-to-word models share the same vocabularies for individual training sets. Following the training approach in \ref{sec:a2w} and \ref{sec:seq2seq}, all end-to-end models trained well. During inference, we observed that the character-based sequence-to-sequence models have confusion when decoding with very long utterances (e.g. 60-120 seconds) so that it performs worse for the \textit{Eval98} test set. The word-based models do not have this issue, however it is theoretically encountered with out-of-vocabulary words.

As shown in Table \ref{tab:end2end_sys}, the end-to-end models trained on the domain-specific sets are also very poor at handling the domain mismatches between training and testing conditions. However, when switching to the multi-domain dataset, all of the end-to-end models behave differently from the hybrid models. As can be seen, the performance of the multi-domain models outperform all the domain-specific models with clear margins. The improvements on the \textit{Hub5'00} and \textit{Libri} test sets clearly indicate that the end-to-end models can exploit the additional training data which come from different domains. This observation also reveals the advantage of the end-to-end approaches over the hybrid approach in multi-domain speech recognition.

In our multi-domain setup, the performance of the multi-domain end-to-end systems are still lacking behind the multi-domain hybrid systems using domain-adapted LMs, but it already surpasses the hybrid systems when the LMs do not match the target test sets, and is at par with the hybrid domain-specific systems.

\section{Domain-conditioned Model}
As previously shown, the end-to-end models can exploit multi-domain data better than the hybrid models. This is probably due to the ability to learn better long-term features. Considering domain identifier (ID) as an abstraction of several long-term features, we investigate an end-to-end model which recognizes the domain of input speech and utilize it during inference. To build such a model, we condition a domain ID for every label sequence in the multi-domain training set before feeding it to a word-based sequence-to-sequence model. This approach is similar to \cite{toshniwal2018multilingual} in which the authors utilize the language ID for building sequence-to-sequence models with a multilingual training set.

Table \ref{tab:dc_sys} shows the WER performance and as well as the accuracy of domain recognition (DAC) of the domain-conditioned model on all the test sets. The DAC is calculated by the percentage of the utterances with correct predictions. Surprisingly, the model recognizes the domain very well on all the test sets while keeping the same WER performance as the end-to-end models in Section \ref{sec:e2e_sys}. 

Since the domains were recognized with a very high accuracy for the test sets having similar conditions with the training set, conditioning the domain-conditioned model with incorrect domain ID lead to worse performance. We additionally evaluated this model on the English part of the Microsoft speech language translation (MSLT) corpus which includes 3,000 recorded utterances over Skype conversations. As the MSLT corpus is an out-of-domain set, we tried to decode the multi-domain LSTM hybrid model using different language models. As shown in Table \ref{tab:dc_sys_mslt}, the best performance is achieved when using \textit{SWB+Fisher} LM which indicates that it is a closer text domain. Interestingly, the domain-conditioned model also achieved better performance when conditioning with the \textit{SWB} domain.

\begin{table}[t]
	\caption{The performance of the domain-conditioned end-to-end models.}
	\label{tab:dc_sys}
	\vspace{-0.2cm}	
	\setlength{\tabcolsep}{4pt}
	\centering
	\begin{tabular}{lcccc}
		\toprule
		& \textbf{Hub5'00} & \textbf{TED} & \textbf{Libri} & \textbf{Eval98} \\
		\midrule
		DAC  & 99.84 & 94.02 & 97.79 & 97.93 \\
		WER  & 18.5 & 10.8 & 8.4 & 11.9 \\
		\bottomrule
	\end{tabular}
	\vspace{-0.2cm}
\end{table}

\begin{table}[t]
	\caption{The performance of the hybrid acoustic models with different LM and the domain-conditioned models with different conditioning domain identifiers on MSLT test set.}
	\label{tab:dc_sys_mslt}
	\vspace{-0.2cm}	
	\setlength{\tabcolsep}{1.5pt}
	\centering
	\begin{tabular}{lcccc}
		\toprule
		\multirow{2}{*}{Hybrid +LM}
		& SWB+Fisher & Cantab & Libri & Multi-Set \\
		& 24.2 & 25.8 & 30.2 & 25.2 \\
		\midrule
		\multirow{2}{*}{Seq2seq +Domain ID}
		& SWB & TED & Libri & Hub4 \\
		& 23.3 & 23.6 & 23.5 & 23.8 \\
		\bottomrule
	\end{tabular}
	\vspace{-0.2cm}
\end{table}

\vspace{-0.2cm}
\section{Conclusions}
We have shown that end-to-end speech recognition models optimized with a sequence-based criterion can generalize better across domains than hybrid acoustic models on a multi-domain setup with diverse domains. We have further shown that a domain identifier can be extracted with high accuracy with a sequence-to-sequence end-to-end model and possibly used to condition the model for performance improvements. These results will lead us to investigating in more detail the question why end-to-end models can learn from diverse acoustic conditions while frame-based hybrid models cannot.
%Future work can explore end-to-end systems on multi-domain training with more different domains.

%\section{Acknowledgements}

\bibliographystyle{IEEEtran}

\bibliography{mybib}

\end{document}